\documentclass[10pt, conference, letterpaper]{IEEEtran}
\IEEEoverridecommandlockouts
\usepackage{cite}
\usepackage{amsmath,amssymb,amsfonts}
\usepackage{algorithmic}
\usepackage{graphicx}
\usepackage{subfigure}
\usepackage{textcomp}
\usepackage{xcolor}
\usepackage{listings}
\usepackage[normalem]{ulem}

\def\BibTeX{{\rm B\kern-.05em{\sc i\kern-.025em b}\kern-.08em
    T\kern-.1667em\lower.7ex\hbox{E}\kern-.125emX}}
\begin{document}

\title{Thwarting Piracy: Anti-debugging Using GPU-assisted Self-healing Codes}

\author{\IEEEauthorblockN{Adhokshaj Mishra}
    \IEEEauthorblockA{
		\textit{Uptycs India Pvt. Ltd.}\\
		\textit{Bengaluru, India}\\
		me@adhokshajmishraonline.in}\\
	\and
	\IEEEauthorblockN{Manjesh 
		K.  Hanawal}
	\IEEEauthorblockA{
	\textit{MLiONS Lab, IEOR, IIT Bombay} \\
		\textit{Mumbai, India}\\
		mhanawal@iitb.ac.in}
}

\maketitle

\begin{abstract}
Software piracy is one of the concerns in the IT sector. Pirates leverage the debugger tools to reverse engineer the logic that verifies the license keys or bypass the entire verification process. Anti-debugging techniques are used to defeat piracy using self-healing codes. However, anti-debugging methods can be defeated when the licensing protections are limited to CPU-based implementation by writing custom codes to deactivate the anti-debugging methods. In the paper, we demonstrate how GPU implementation can prevent pirates from deactivating the anti-debugging methods by using the limitations of debugging on GPU. Generally, GPUs do not support debugging directly on the hardware, and therefore all the debugging is limited to CPU-based emulation. Also, a process running on CPU generally does not have any visibility on codes running on GPU, which comes as an added benefit for our work. We provide an implementation on GPU to show the feasibility of our method. As GPUs are getting widespread with the raise in popularity of gaming software, our technique provides a method to protect against piracy. Our method thwarts any attempts to bypass the license verification step thus offering a better anti-piracy mechanism.
\end{abstract}


\section{Introduction}
Anti-debugging techniques are specially crafted chunks of code that involve one or more methods to detect, and possibly prevent debugging attempts on the target process. Most of the time, these codes are integrated into the final binary blob of the program they are trying to protect; however "out of process" anti-debugging is also possible by employing system-level hooking in user mode, or kernel mode.
 \begin{figure*}[!t]
    \centering
    \includegraphics[scale=.35]{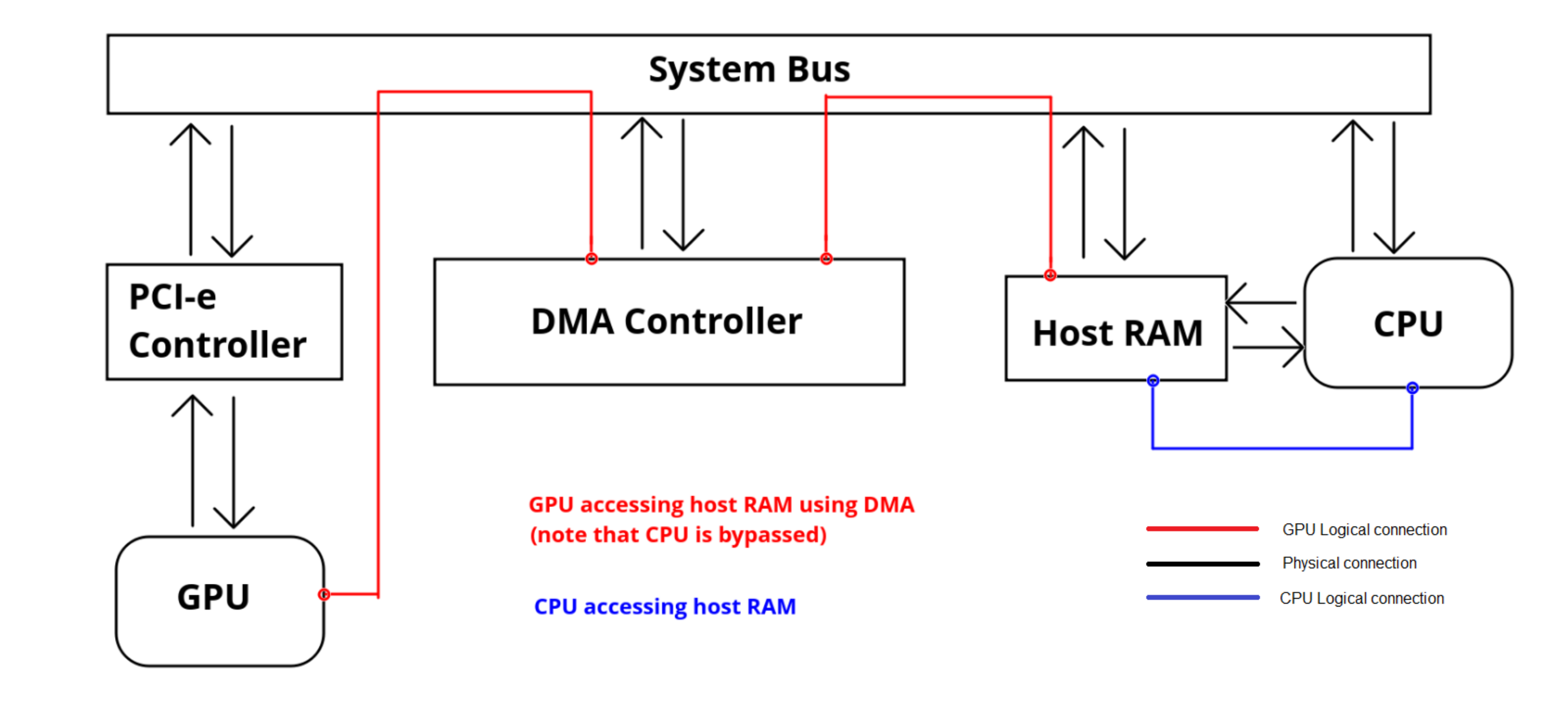}
    \caption{Memory Access Diagram}
    \label{fig:MemoryAccess}
\end{figure*}

Since most of the debuggers rely on modification of debuggee process to large extent (e.g. break-points are set by overwriting instructions, changing permissions, altering error handler chains, etc.), anti-debugging techniques mostly rely on detecting such modifications and possibly reverting them to their original state. Since reverting to the original state requires modification of our own process, we commonly see some variation of self-modifying codes being used for such purposes.

Self-healing codes are a special case of self-modifying codes, which have the ability to detect any modifications in their code and revert it to its original state before executing it. Although these techniques can restore sufficiently large modifications, in practice these are kept limited to only critical parts of the code. These are often used to subvert debugging techniques allowing the protected processes to evade from the eyes of debugger tools (e.g. process not stopping on a break-point).

Anti-debugging techniques generally can be grouped in the following categories though not all techniques are applicable to all platforms:\\
\noindent
 \textbf{1. Timing and Latency Analysis}: These techniques rely on the difference in time taken to run a known calibrated code. Debugging tools generally make the underlying process run a bit slower due to their invasive nature.\\
 \noindent
\textbf{2. Process Detection}: These techniques rely on detecting the presence of known debugging and related tools. If such tools are running on a system, the chances of the process being under watch is fairly high.\\
 \noindent
 \textbf{3. Memory Analysis}: Many debugging tools tend to alter memory maps in different ways (different parts of code and data being loaded in slightly different locations, differences in stack, heap, etc), which can be used to detect debugging in such cases.\\
\noindent
\textbf{4. Break-point Detection}: These techniques rely on the fact that setting a software break-point alters machine code in memory. The program scans its own memory to search machine code for software break-point (e.g. 0xCC on x86 and AMD64) in code regions.\\
\noindent
\textbf{5. Patching Detection}: These techniques are one step ahead of techniques described in (4), and these are able to detect arbitrary patching (machine code modification) in process memory.\\
\noindent \textbf{6. Monitoring Debugger APIs}: These techniques rely on the fact that only one process can act as a debugger for another given process at a time. In other variations, debugger APIs exposed by the platform are hooked and monitored globally to detect debugging attempts (e.g. DebugActiveProcess(...) / WaitForDebugEvent(...) on Windows, ptrace on Linux).\\
\noindent \textbf{7. Monitoring Exception Handlers}: Many times when a debugger is attached to a process, exceptions are trapped and handled by the debugger without passing the exception back to the application for continued execution. Occasionally these exceptions can even crash or terminate a process when run under a debugger and be handled gracefully when running without a debugger attached. These discrepancies can be used to detect debugging attempts. \\
\noindent
\textbf{8. Blocking Debuggers}: These techniques rely on either blocking Debugger APIs (using self-debugging codes), or removing debugger-related artifacts like break-points or in-memory patching from protected code.

Software piracy is one of the biggest problems facing the IT industry. The act of piracy revolves around figuring out the licensing method used by the target software, reverse engineering it, or making the software bypass license checks. Debuggers are the go-to tools for this purpose. To defeat this, anti-debugging codes are commonly used for anti-piracy, IP protection, and digital rights management. These codes generally make it hard for pirates to figure out licensing or IP/digital rights protection logic, making it infeasible or more costly to bypass license checks than paying for a legitimate copy of the software. In certain cases, such techniques may also make it almost impossible to pirate content without using special hardware. For example, video streaming platforms often rely on High-Bandwidth Digital Content Protection (HDPC) copy-protection which encrypts the signals between computer and display, thereby making it almost impossible to copy the video content in order to pirate it.  Most of the software licensing protections are limited to CPU-based implementations which can be defeated by writing custom tools (e.g. custom key management server implementations used to be really popular to illegally activate Windows OS installations \cite{WindowsLink}, or by employing techniques specific to anti-debugging techniques being used. We will cover a survey of common CPU-based anti-debugging techniques and their shortcomings in upcoming sections.

Since hardware components like GPU are becoming commonplace, and are coming with general compute capabilities, even in consumer-grade versions, these can also be potentially used to implement various anti-debugging techniques almost entirely on GPU. This can be done even more easily for software that needs a dedicated GPU to function properly, like image editing, video editing, gaming, rendering software, and CAD software. Since the GPU stack generally does not have very detailed debugging, analysis, and instrumentation capabilities  similar to CPU stack, these techniques can raise the bar very significantly. For example, NVIDIA toolkit provides GPU code debugging using GPU emulation on many cards but not on the GPU card itself.

In the past, pirated copies of many licensed software were distributed widely (e.g., Photoshop). This was possible as they implemented CPU-based piracy protection mechanisms which could be bypassed by disabling anti-debugging features.  To overcome these issues, software license verification is moved to the cloud, but this needs an additional layer of authentication. However, for many software, especially, gaming software, it is preferable that the license verification is performed locally. Our method provides a mechanism for such verification without requiring additional layer authentication while overcoming the issues faced in CPU-based license verification. 





In this paper, we present another technique that involves running anti-debugging code on GPU and monitoring the process on the CPU to protect it from getting debugged. Our method exploits the fact that
CPU hardware breakpoints do not have visibility in GPU code. We run the anti-debugging code to remove the breakpoints in GPU. The GPU performs the task by using Direct Memory Access (DMA) on which the CPU does not have visibility. Due to the lack of visibility, it is not possible to prevent breakpoints erasures in the CPU. Once breakpoints are erased, the CPU continues the normal execution of the verification process. Thus our method thwarts any attempt to deactivate the anti-debugging method and achieve piracy. 

Figure \ref{fig:MemoryAccess} gives the memory access architecture. Note that only the CPU has direct access to the host memory through  CPU logical connections. GPU access host memory only through the DMA controller. Hence CPU does not have visibility of the memory access activities of the GPU. 

The paper is organized as follows: In Section \ref{sec:AntiDebugCPU} we discuss various anti-debugging methods in CPU and discuss their limitations in \ref{sec:AntiDebugLimit}. In Section \ref{sec:GPUAntiDebug} demonstrates how GPU can assist in improving the ant-debugging capabilities of the CPUs and provide develop a robust mechanism for thwarting piracy.  Section \ref{sec:Conclusion} gives conclusion. 

\subsection{Related Work}
Piracy has been one of the long concerns of the software industry and several works have addressed the issue. \cite{FSE2000_SoftwareEngineeringSecurity} give an early roadmap of the security concerns. Since then several methods are proposed 
various methods to improve software piracy, like smartcard based method \cite{CEC2002_EnhanceSmartCard}, code obfuscation \cite{ACMComputingServey2016_ProtectingSoftware}, and through digital rights management systems \cite{CEC2005_SoftwarePrivacyPrevention}. 

Recently studies have been undertaken to understand how personality traits \cite{MySec2015} and economic conditions \cite{ICAITI2019_SoftwarePrivacyFactorProfiling} influence software privacy. Various methods are also proposed to detect attempts at software privacy. 
\cite{IMISUC2014_PreventSoftwarePrivacy} proposed a method to detect distribution of Windows Executable Programs via bit torrent and \cite{ICCKE2016_MetaSPD} proposed MetaSPD that performs metamorphic analysis to automatically detect pirated software copies. 

Anti-bugging is a feature used to prevent reverse engineering of code through debugging process and has been extensively used to achieve anti-piracy  \cite{IEEESecurityandPrivacy2007_SoftwareProtection}. However, pirates have been able to evade it using techniques similar to that used by malware  \cite{DSN2008_AntiVirtualizationAntiDebugging}, \cite{COMSNETS2021_EavdingMalware}.

Our work proposed to enhance the anti-bugging feature by using GPU-assisted self-healing code. To the best of our knowledge, GPU-assisted anti-piracy techniques are not studied in the literature. All the missing details in this paper are given in the extended version of the paper \cite{TechReport}.    

\section{Anti-debugging on CPU}
In this section we will discuss various modes to detect if a debugger is attached to the code and how to use it in anti-debugging. We begin with a discussion of types of breakpoints. 

\subsection{Type of breakpoints}
A breakpoint is an intentional “pause” in the normal execution of a program, generally used to inspect the internals of said process in more detail. This is the most used feature of any debugger. On x86 CPU, there are two types of breakpoints: hardware breakpoints and software breakpoints. While they overlap but are not exactly the same. \\

\noindent
\textbf{Software breakpoint:} In most of debugging cases software breakpoints are used, which do not need any special hardware support. These are implemented using the interrupt mechanism provided by CPU. On x86 interrupt number 03 is used to implement a software breakpoint by convention. When a breakpoint is set, the debugger overwrites the target address where we want to put the break-point with INT 03H (0xCC in hex). When this instruction gets executed, the debugger gets the control back from the target process and can inspect its state (registers, memory, etc). To resume the execution, the debugger will silently remove the break-point, execute the instruction, and set the break-point again before letting the process resume until it terminates or breaks. Generally, we can set any number of software breakpoints; however these cannot be set on non-code addresses, i.e., these can break the program only when target address content is executed; but not if the address is read from or written to.

\noindent
\textbf{Hardware breakpoint}: Hardware breakpoints, on the other hand, are much more powerful and flexible than software breakpoints. These can be set to break not only on execution but also on memory access (read and write both), I/O port access, etc. These debuggers are set by writing into special “debug registers” which are largely platform specific. Not all platforms will have support for hardware breakpoints.

In x86 architecture, the debugger uses a set of Debug Registers in order to apply hardware breakpoints. There exits $8$ debug registers to control the debugging procedure, ranging from DR0 to DR7. These registers are not accessible from ring3 privileges but only accessible from Current Privilege Levels, ring0 (CPL0). Thus, an attempt to read or write the debug registers when executing at a privilege level other than CPL0 causes a general protection fault. The debug registers allow the debugger to interrupt program execution and transfer the control to it when accessing memory to read or write.

x86 has the following debug registers:
\begin{enumerate}
    \item DR0-DR3: Linear break-point address 0-3. The stored address can be the same as the physical address or it needs to be translated to the physical address.
    \item DR4-DR5: Reserved. Not defined by Intel
    \item DR6: Break-point status, which indicates which break-point is activated.
    \item DR7: Break-point control, which defines the break-point activation mode by the access modes: \textit{read}, \textit{write}, or \textit{execute}.
\end{enumerate}

\noindent
Some debuggers can also feature other types of breakpoints:

\noindent
\textbf{Memory breakpoint}: Memory breakpoints are implemented by a debugger using guard pages. when a page of memory is marked as PAGE{\_}GUARD and is accessed, a STATUS{\_}GUARD{\_}PAGE{\_}VIOLATION exception is raised, which can then be handled by the debugger. However, such implementations can be Windows-specific.

\subsection{Detecting Software Breakpoint}
Since we know that software breakpoints are set by overwriting 0xCC at the first byte of the instruction, we can easily check for such breakpoints in our code:
\begin{enumerate}
    \item Find offsets of all instructions in the target function, starting from the location of the first instruction.
    \item Find where our target function, or any chunk of code, is located in memory
    \item Read one byte from all offsets
    \item If any byte is 0xCC, a break-point has been set 
\end{enumerate}
A simple implementation code in C++ is as follows:

\begin{lstlisting}[frame=single]
bool isBreakpointPresent(
    const unsigned char *func, 
    const std::vector<unsigned int>& offsets)
{
    bool result = false;
    for (auto &i : offsets)
    {
        if (*(func + i) == 0xCC)
        {
            result = true;
            break;
        }
    }
    return result;
}
\end{lstlisting}

For the above technique, an analyst can reverse engineer it, find its location in a compiled binary (or memory address at run-time), and modify it so that it always returns the value which is expected by the rest of the code. In our case, it will be boolean value "false", because this is what we are returning when no breakpoint is present. Once this function is modified, the rest of the code will not be able to detect the presence of breakpoints and will continue working normally, which will assist the analyst further in reverse engineering the rest of the code. 

To deal with this, one can try to detect function patching, which can be done by the following two methods:

\begin{enumerate}
    \item By matching monitored code byte by byte with a known good copy of code
    \item By calculating a checksum of code, and comparing it with known good checksum
\end{enumerate}

The first one is simple to implement but inferior to the other option as multiple copies of the same code have to be maintained, which increases size. For sake of simple implementation of technique (2), we can use CRC and compare it with known good value. A reference implementation is given in \cite{TechReport}:

CRC-based implementation can also be bypassed by modifying protected code as well as hard-coded correct checksum values. This is mitigated by not hard coding the checksum, but using it as a parameter for some other code instead. Using checksum to decrypt some other stuff, or using it in some jump/lookup table are some possible methods.

\subsection{Detecting Hardware Break-point}

Detecting hardware breakpoint involves OS-specific techniques, as different operating systems expose underlying hardware details in different ways. On Windows, this can be done using \texttt{GetThreadContext} API to get thread context for a given thread, and then inspect values of debug registers. On Linux/BSD, this can be achieved by installing a system-wide hook for \texttt{ptrace} system call, and monitoring parameters for \texttt{PT{\_}GETDBREGS}, \texttt{PT{\_}SETDBREGS}, \texttt{PTRACE{\_}PEEKUSER} and \texttt{PTRACE{\_}POKEUSER} trace calls.

We note that unlike the \texttt{GetThreadContext} example, the above example cannot be used by the process itself. One has to monitor/protect the desired process from the outside process.

\subsection{Evading Software Break-point}

Since we already know how a software break-point works, we can create a method to evade such break-points as follows:

\begin{enumerate}
    \item Check if a software break-point is present on the target code.
    \item Find the offsets (or locations) where a break-point is applied.
    \item Disable memory protection on the memory block containing the target code.
    \item Restore original bytes at affected offsets
    \item Restore original memory protection
\end{enumerate}

This method can be trivially enhanced to restore code in case of function patching. A reference implementation for the above pseudocode and its enhancements are given in \cite{TechReport}.

\subsection{Evading Hardware Break-point}
Just like the detection of hardware breakpoints, their evasion is also tightly coupled to the specific platforms.

First, we present a possible algorithm, and its reference implementation for Windows, which removes hardware break-point using custom installed Structured Exception Handler (called SEH henceforth). This mechanism is commonly seen in Windows malware. which is done using the following routines:\\

\noindent
\textbf{Routine ClrHwBpHandler:}
This routine works in the following steps.
\begin{enumerate}
    \item Zero out a suitable register (we use EAX)
    \item Find address of CONTEXT structure in the stack (this is at an offset of 0xC from the current position in the stack when our routine is triggered from SEH chain
    \item Reset values of DR0-DR3, DR6, and DR7 to 0, by writing at proper offsets from the beginning of CONTEXT structure.
    \item 
    Once our handler completes, OS will try to resume the process from the same instruction where the fault occurred. Since we do not want to repeat that instruction anymore, we will modify the instruction pointer, and the process will resume from whatever address/offset we provide in the instruction pointer. Since CONTEXT structure takes offset in the instruction pointer, we have to put the size of the instruction (as the number of bytes to skip) at  Extended Instruction Pointer (EIP) in CONTEXT structure.
    \item Return from the routine. After this, the system will resume execution from our specified EIP.
\end{enumerate}

\noindent
\textbf{Routine ClearHardwareBreakpoints:}
This routine works in the following steps.
\begin{enumerate}
    \item Setup routine ClrHwBpHandler as SEH handler
    \item Perform some operation that triggers a fault. We are using divide by 0.
    \item Add the rest of the code as usual. This is the code that is to be protected from debugging.
\end{enumerate}

\begin{lstlisting} [frame=single]
; routine has to be declared before we can refer to it
ClrHwBpHandler proto
 .safeseh ClrHwBpHandler

; routine ClearHardwareBreakpoints starts here
ClearHardwareBreakpoints proc
    ; step (1) starts here
    assume fs:nothing
    push offset ClrHwBpHandler
    push fs:[0]
    mov dword ptr fs:[0], esp ; Setup SEH
    
    ; step (2) starts here
    xor eax, eax
    div eax ; Cause an exception
    
    ; step (3) starts here
    pop dword ptr fs:[0] ; Execution continues here
    add esp, 4
    ret
ClearHardwareBreakpoints endp

; routine ClearHardwareBreakpoints ends here

; routine ClrHwBpHandler starts here
ClrHwBpHandler proc 
    ; step (1) starts here
    xor eax, eax
    
    ; step (2) starts here
    mov ecx, [esp + 0ch] ; This is a CONTEXT structure on the stack
    
    ; step (3) starts here
    mov dword ptr [ecx + 04h], eax ; Dr0
    mov dword ptr [ecx + 08h], eax ; Dr1
    mov dword ptr [ecx + 0ch], eax ; Dr2
    mov dword ptr [ecx + 10h], eax ; Dr3
    mov dword ptr [ecx + 14h], eax ; Dr6
    mov dword ptr [ecx + 18h], eax ; Dr7
    
    ; step (4) starts here
    add dword ptr [ecx + 0b8h], 2 ; We add 2 to EIP to skip the div eax
    
    ; step (5) starts here
    ret
ClrHwBpHandler endp

; routine ClrHwBpHandler ends here
\end{lstlisting}

On Linux (and BSD platforms), hardware registers cannot be cleared by the protected process itself. Just like detection, these have to be done via a different process, or via a system-level hook. A sample implementation of hardware breakpoint removal via different processes is given below:

\begin{lstlisting} [frame=single]
#define DR_OFFSET(x) (user->u_debugreg + x)

unsigned long long setDebugRegister(
    const user* user, 
    const pid_t pid, 
    unsigned char index, unsigned long long value)
{
    unsigned long long result = 0;
    result = ptrace(PTRACE_PEEKUSER, pid, user->u_debugreg[index], 0);
    
    ptrace(PTRACE_POKEUSER, pid, user->u_debugreg[index], &value);
    
    return result;
}

void RemoveHardwareBreakpointPresent(
    const user* user, 
    const pid_t pid)
{
    unsigned long long dr[4];
    
    for (int i = 0; i < 4; ++i)
    {
        dr[i] = setDebugRegister(user, pid, i, 0);
    }
}
\end{lstlisting}

\subsection{Evading Memory Breakpoint}

Since these are implemented in different debuggers in different ways, we need to figure out the specific implementation used by debuggers that we are trying to protect from. Assuming that a debugger is using the implementation given earlier, we can detect it by using the following logic:

\begin{enumerate}
    \item Allocate a dynamic buffer
    \item Write machine code for \texttt{RET} in the beginning of the buffer
    \item Mark the page as a guard page
    \item Push a return address (starting address of code which will be run on successful return) on the stack
    \item Make an unconditional jump to guard page
    \item If the code at our custom return address gets executed, it means that exception was caught by a debugger. Therefore, our process is being debugged. Once detected,  we evade the anti-bugging process by either exiting the process or  changing the behavior.
\end{enumerate}

\label{sec:AntiDebugCPU}

\section{Limitations of anti-debugging on CPU}
All the techniques that we have covered so far run entirely on CPU, and therefore can eventually be circumvented by an analyst. For sake of completeness, we document different techniques and their detection and evasion mechanisms below:

\begin{enumerate}
    \item \textbf{Detecting software breakpoint}: This can be detected by setting up hardware breakpoints for memory read on suspected memory addresses.
    \item \textbf{Detecting hardware breakpoint}: This can be detected by monitoring calls to \texttt{GetThreadContext()} API in Windows and \texttt{ptrace()} system call in (Linux / BSD) at system level.
    \item \textbf{Evading software breakpoint}: This can be detected by setting up hardware breakpoints for memory writing on suspected memory addresses. This can also be detected by comparing process memory snapshots taken at different timestamps to see if there are any changes in code in memory.
    \item \textbf{Evading hardware breakpoint}: This can be detected by monitoring \texttt{SetThreadContext()} API and SEH handlers in process in Windows, or \texttt{ptrace()} system call at system level in Linux / BSD.
\end{enumerate}
\label{sec:AntiDebugLimit}

\section{GPU-assisted Anti-Debugging}
To tackle the limitations of anti-debugging techniques running on CPU, we propose a new technique, which involves running part of the code on GPU. It gives us the following benefits:

\begin{enumerate}
    \item Code running on GPU is normally visible in system. In very special cases it can be visibility, but this visibility is limited, like run information cannot be extracted.
    \item Due to the popularity of toolkits like CUDA/OpenCL/AMD APP, programming a GPU is almost as easy as writing a program for CPU.
    \item If a process invokes a code on GPU, it looks like a vague instruction from the process's point of view, i.e., it invokes and then waits for it to complete, while the GPU code is running on GPU.
    \item Tools to debug and reverse engineer GPU-specific codes are not as common as their CPU counterparts. To complicate it more, many tools either simulate the hardware instead of debugging on real hardware or are specific to certain vendors and/or models of GPUs.
\end{enumerate}


Since most of the GPUs are connected to the motherboard and CPU using PCI express bus (a high-speed serial computer expansion bus providing a common interface for general hardware like GPU, storage adapters, network adapters, etc. It also provides advanced error detection and reporting, hot-swapping, and I/O virtualization. We can set up "direct memory access" (called DMA hereafter) between GPU and host memory (RAM). The DMA setup allows us to read/write from/to host memory without having to involve a CPU. In common implementations, CPU and other peripherals are connected to a common bus, and many of these devices can take control of the bus, to read/write into some other hardware via memory mapped I/O or host memory itself. Since the DMA does not involve CPU, it can be used to evade certain hardware-enabled monitoring features.

We also end up having to deal with the following disadvantages:

\begin{enumerate}
    \item We cannot use these techniques where a dedicated GPU is not present, or is not available for some reason (e.g. we are on host, but GPU is connected to some VM via PCI pass-through)
    \item Not all GPU programming toolkits are equal (different set of features, different support for various hardware and their capabilities). It may limit us to some specific toolkits, which can further limit us to hardware from specific vendor and/or specific model / series.
\end{enumerate}

In this paper, we will convert the method of software breakpoint detection and its corresponding removal method to run on GPU. For this, we will follow the logic given below:

\textbf{On GPU}
\begin{enumerate}
    \item Setup GPU to access protected function on host from GPU
    \item Call \texttt{isBreakpointPresent()} on GPU, and wait for it to complete.
    \item Copy the result from GPU to host, and check it to see if there is any breakpoint or not.
    \item Find memory pages which are hosting code corresponding to protected function
    \item Change memory permission on memory pages (from the previous step) to read, write and execute.
    \item Set up DMA between GPU and host memory (RAM), and map the pages from step (6) to GPU memory.
    \item Call \texttt{RemoveBreakpoint()} on GPU, and wait for it to complete.
\end{enumerate}

For reference implementation and its testing, we have used the following setup:

\begin{itemize}
    \item \textbf{Host OS}: Arch Linux x64 (Linux kernel: 5.18.7-arch1-1)
    \item \textbf{Host CPU}: Intel Core i7-6700HQ CPU
    \item \textbf{GPU}: NVIDIA Corporation GM107M [GeForce GTX 960M]
    \item \textbf{GPU Driver}: NVIDIA 515.48.07-13
    \item \textbf{GPU Programming Toolkit}: NVIDIA CUDA 11.7.0-2
\end{itemize}

We note that although we implemented and tested this technique on Linux + NVIDIA combination, this can be ported easily to non-Linux platforms as well for NVIDIA GPUs. For GPUs from other vendors, corresponding toolkits may be used.

The reference implementation is as below:

\begin{lstlisting} [frame=single]
#include <stdio.h>
#include <iostream>
#include <sys/mman.h>
#include <unistd.h>

// step (1) starts here
__global__ void isBreakpointPresent(
    unsigned char *func, 
    int *result)
{
    unsigned int offsets[] = 
        {0, 1, 4, 8, 15, 17, 24, 27, 34, 
        37, 42, 49, 52, 55, 60, 64, 68, 
        70, 71, 72, 73};
        
    bool tmp = false;

    for (int i = 0; i <= 20; ++i)
    {
        if (*(func + offsets[i]) == 0xCC) {
            tmp = true;
        }
    }

    if (tmp)
        *result = 1;
    else
        *result = 0;
}
// step (1) ends here

// step (2) starts here
__global__ void removeBreakpoint(unsigned char *func)
{
    unsigned int offsets[] = 
        {0, 1, 4, 8, 15, 17, 24, 27, 34, 
        37, 42, 49, 52, 55, 60, 64, 68, 
        70, 71, 72, 73};
        
    unsigned char original_bytes[] = 
        {0x55, 0x48, 0x48, 0xc7, 0xeb, 
         0x48, 0x48, 0x48, 0x48, 0xe8, 
         0x48, 0x48, 0x48, 0xe8, 0x83, 
         0x83, 0x7e, 0x90, 0x90, 0xc9, 
         0xc3};

    for (int i = 0; i <= 20; ++i)
    {
        if (*(func + offsets[i]) != original_bytes[i]) {
            *(func + offsets[i]) = original_bytes[i];
        }
    }
}
// step (2) ends here

// protected function, referenced in step (3)
void secret()
{
    for (int i = 0; i < 10; ++i)
    {
        std::cout << "Try a breakpoint at secret()" << std::endl;
    }
}

int main(void)
{
    int *result;
    int h_result;
    
    // initialize CUDA
    cudaMalloc(&result, sizeof(int));

    // change memory permission to 
    // read/write/execute. This is needed 
    // to change memory contents from GPU
    
    // step (6) starts here
    long pagesize = sysconf(_SC_PAGESIZE);
    unsigned long page_start = (unsigned long)secret & ~(pagesize - 1);
    // step (6) ends here
    
    // step (7) starts here
    if (mprotect((void*)page_start, pagesize, PROT_READ | PROT_WRITE | PROT_EXEC) != 0)
    {
        std::cerr << "mprotect() failed" << std::endl;
    }
    // step (7) ends here

    // setup GPU to access host memory 
    // (system RAM) from GPU via direct 
    // memory access
    
    // step (8) starts here
    cudaError err = cudaHostRegister((void*)secret, 75, cudaHostRegisterMapped);
    if (err != cudaSuccess) {
        fprintf(stderr, "Host register failed for function with error %d\n", err);
    }
    // step (8) ends here

    // find address of protected function 
    // on system RAM
    unsigned char* func = (unsigned char*)secret;

    // and convert it to address from GPU
    cudaHostGetDevicePointer(&func, (void*)secret, 0);

    // check if breakpoint is present.
    // this runs on GPU
    
    // step (4) starts here
    isBreakpointPresent<<<1,1>>>(func, result);

    cudaDeviceSynchronize();
    err = cudaGetLastError();
    if(err!=cudaSuccess)
    {
        fprintf(stderr,"ERROR: %s\n", cudaGetErrorString(err) );
        exit(-1);
    }
    // step (4) ends here

    // copy output of breakpoint check to 
    // variable in host memory
    
    // step (5) starts here
    cudaMemcpy(&h_result, result, sizeof(int), cudaMemcpyDeviceToHost);

    cudaDeviceSynchronize();
    err = cudaGetLastError();
    if(err!=cudaSuccess)
    {
        fprintf(stderr,"ERROR: %s\n", cudaGetErrorString(err) );
        exit(-1);
    }

    cudaFree(result);
    // step (5) ends here

    if (h_result == 1) {
        std::cerr << "secret() has been hooked" << std::endl;
        
        // remove breakpoint, breakpoint 
        // is present the following line 
        // will run on GPU
        
        // step (9) starts here
        removeBreakpoint<<<1,1>>>(func);
        cudaDeviceSynchronize();
        // step (9) ends here

        func = (unsigned char*)secret;
        if (*func == 0xCC)
            std::cerr << "secret() is hooked" << std::endl;
        else
            std::cerr << "hook at secret() has been removed" << std::endl;
        secret();
    }
    else
    {
        std::cerr << "secret() has not been hooked" << std::endl;
        secret();
    }
    
    // tear down everything
    cudaHostUnregister((void*)secret);
}
\end{lstlisting}

Note that the above code can be  extended to detect patching, and restore protected function(s) to original state(s).

If we run the above code inside debugger and setup a breakpoint manually on the \texttt{secret()} function, then set a hardware watch-point to detect when the breakpoint is removed; we will see that hardware watch-point \textbf{does not trigger}:


\begin{lstlisting} [frame=single]
$ gdb -q ./cuda
Reading symbols from ./cuda...
(gdb) break main
Breakpoint 1 at 0xc204: file ../main.cu, line 55.
(gdb) run
Breakpoint 1, main () at ../main.cu:55
55	{
(gdb) disassemble secret
Dump of assembler code for function _Z6secretv:
   0x00005555555601b2 <+0>: push %rbp
   ...
End of assembler dump.
(gdb) set *((char*)0x00005555555601b2) = 0xCC
(gdb) watch *0x00005555555601b2
Hardware watchpoint 2: *0x00005555555601b2
(gdb) continue 
Continuing.
secret() has been hooked
hook at secret() has been removed
Try a breakpoint at secret()
...
[Inferior 1 (process 11907) exited normally]
\end{lstlisting}

Please note that lines starting with \$ are shell prompt, and lines starting with (gdb) are debugger prompt.

In the above output, we have done the following:

\begin{itemize}
    \item \textbf{gdb -q ./cuda}: cuda is the compiled binary here, which we are starting to load via debugger (GNU debugger in this case). The argument '-q' is passed to prevent debugger from printing elaborate messages.
    \item \textbf{(gdb) break main}: We are setting a software breakpoint on main() function, which is entry point of our PoC code. We are doing this because we want actual addresses where out functions get loaded at runtime.
    \item \textbf{(gdb) run}: We ask the debugger to run the given input program (named cuda), and wait for any "debug event" like hitting a breakpoint. Please note that debugger stops execution of input program as soon as breakpoint is hit, and we get another debugger prompt.
    \item \textbf{(gdb) disassemble secret}: We ask the debugger to print assembly listing of function named secret(). In output, debugger prints starting addresses, as well as assembly instructions line by line. We have stripped the listing to keep it short.
    \item \textbf{(gdb) set *((char*)0x00005555555601b2) = 0xCC }: Change first byte at address to 0xCC, which is machine code for software breakpoint. This effectively sets a software breakpoint on secret() function.
    \item \textbf{(gdb) watch *0x00005555555601b2}: Setup a hardware breakpoint on given address. This breakpoint will trigger if a write is performed on given address. For breaking on read, we need to use rwatch instead of watch.
    \item \textbf{(gdb) continue}: Continue the execution, until some "debug event" happens.
\end{itemize}

After the last step, we see that:

\begin{itemize}
    \item Input program continues execution
    \item Breakpoint on secret() is detected.
    \item Breakpoint on secret() is removed, but no hardware breakpoint is triggered.
    \item Secret() is executed.
    \item Input program completes execution, and exits.
\end{itemize}
\label{sec:GPUAntiDebug}

\section{Conclusion}
In this paper, we demonstrated feasibility of a anti-debugging technique to thwart piracy. It relies on running break-point detection and removal code on GPU, and protects the target function by modifying its contents in host RAM via DMA between GPU and host memory; thereby defeating the hardware watch-point mechanism provided by host CPU to monitor changes in host memory. We have demonstrated the aforementioned technique on a Linux x64 machine having NVIDIA GPU using the NVIDIA CUDA toolkit.
\label{sec:Conclusion}



\bibliographystyle{IEEEtran}
\bibliography{main.bib}


\end{document}